\documentclass[10pt,preprint]{aastex}

\shorttitle{The Accuracy of Monochromatic Extrapolated Infrared Luminosities}
\shortauthors{Lin et al.}

\usepackage{natbib}
\usepackage{multirow}
\usepackage{booktabs}
\usepackage{bookmark}

\newcommand{\um}{$\mathrm{\mu m}$}
\begin{document}

\title{How accurate are infrared luminosities from monochromatic photometric extrapolation?}

\author{Zesen Lin\altaffilmark{1}, Guanwen Fang\altaffilmark{2,3}, and Xu Kong\altaffilmark{1}}

\altaffiltext{1}{CAS Key Laboratory for Research in Galaxies and Cosmology, Department of Astronomy, University of Science and Technology of China, Hefei, Anhui 230026, China; zesenlin@mail.ustc.edu.cn, xkong@ustc.edu.cn}
\altaffiltext{2}{Institute for Astronomy and History of Science and Technology, Dali University, Dali 671003, China; wen@mail.ustc.edu.cn}
\altaffiltext{3}{Zesen Lin and Guanwen Fang contributed equally to this work.}

\begin{abstract}
Template-based extrapolations from only one photometric band can be a cost-effective method to estimate the total infrared (IR) luminosities ($L_{\mathrm{IR}}$) of galaxies. By utilizing multi-wavelength data that covers across 0.35--500\,$\mathrm{\mu m}$ in GOODS-North and GOODS-South fields, we investigate the accuracy of this monochromatic extrapolated $L_{\mathrm{IR}}$ based on three IR spectral energy distribution (SED) templates (\citealt[CE01]{Chary2001}; \citealt[DH02]{Dale2002}; \citealt[W08]{Wuyts2008a}) out to $z\sim 3.5$. We find that the CE01 template provides the best estimate of $L_{\mathrm{IR}}$ in {\it Herschel}/PACS bands, while the DH02 template performs best in {\it Herschel}/SPIRE bands. To estimate $L_{\mathrm{IR}}$, we suggest that extrapolations from the available longest wavelength PACS band based on the CE01 template can be a good estimator. Moreover, if PACS measurement is unavailable, extrapolations from SPIRE observations but based on the \cite{Dale2002} template can also provide a statistically unbiased estimate for galaxies at $z\lesssim 2$. The emission of rest-frame 10--100\,$\mathrm{\mu m}$ range of IR SED can be well described by all the three templates, but only the DH02 template shows nearly unbiased estimate of the emission of the rest-frame submillimeter part.
\end{abstract}


\section{Introduction}
\label{sect:intro}

The infrared (IR) sky is dominated by emission from star-forming galaxies (\citealt{Lagache2005,Viero2009,Viero2013}). Massive, young stars emit a large amount of ultraviolet (UV) radiation, which is absorbed by surrounding dust grains and then re-emitted at IR wavelength. Physical properties, such as dust components, dust temperature, star formation rate (SFR), can be decoded from the IR spectral energy distributions (SEDs) of galaxies. The total bolometric infrared luminosity $L_{\mathrm{IR}}$, which is simply an integration of the SED in a wavelength range that most often in 8--1000\,$\mathrm{\mu m}$ (e.g., \citealt{Kennicutt1998}), traces the total energy absorbed by dust. Hence $L_{\mathrm{IR}}$ can be used to estimate the obscured SFR (\citealt{Kennicutt1998,Kennicutt2012}), although non-star formation driven heating, like old stellar populations or active galactic nucleus (AGNs) heating, may have a non-negligible contribution in some galaxies.

The most reliable method to estimate $L_{\mathrm{IR}}$ is deriving it from observations that sampling nearly the whole IR wavelength by directly integrating or multi-wavelength fitting of empirical templates (e.g., CE01, DH02) or physical dust models (e.g., \citealt{SiebenmorgenKruegel2007}). However, for the majority of galaxies we concerned, observations in IR bands are not enough to allow a direct integration even a reliable SED fitting. Using photometric data as less as possible to extract information about $L_{\mathrm{IR}}$ would be a cost-effective way to solve this problem. \cite{Sanders1996} provided an equation that using four {\it Infrared Astronomical Satellite} ({\it IRAS}) flux densities at 12, 25, 60 and 100\,$\mathrm{\mu m}$ to estimate $L_{\mathrm{IR}}$. DH02 derived two similar relations based on three {\it IRAS} bands (25, 60 and 100\,$\mathrm{\mu m}$) and three {\it Spitzer}/Multiband Imaging Photometer for {\it Spitzer} (MIPS) bands (24, 70 and 160\,$\mathrm{\mu m}$) but for the 3--1100\,$\mathrm{\mu m}$ bolometric luminosity. Recently, \cite{Dale2014} updated the previous results and added possible contribution from AGN, derived one {\it IRAS}-based and two {\it Spitzer}-based relations to estimate the total luminosity over 5--1100\,$\mathrm{\mu m}$ for a range of mid-infrared AGN fractions. Moreover, \cite{Boquien2010} provided detailed relations to estimate $L_{\mathrm{IR}}$ from just one or two {\it Spitzer} bands, especially from 8 and 24\,$\mathrm{\mu m}$ bands, calibrated by local galaxy samples. {\it Herschel}-based (from 70, 100, 160 and 250\,$\mathrm{\mu m}$ bands) relations were also developed by \cite{Galametz2013} using observations of local galaxies.

We note that most of the empirical relations between $L_{\mathrm{IR}}$ and observed flux densities provided by above works require more than one photometric observations, while a few relations only need observation in one band but calibrated by local galaxies. One typical monochromatic method that easily generalizes to high-redshift galaxies is extrapolating $L_{\mathrm{IR}}$ based on IR SED templates. Following this method, \cite{Elbaz2010} have utilized observations from {\it Herschel} to  check the self-consistency of the used IR SED templates. However, it is still necessary to carefully investigate the accuracy of this method by applying a more actual $L_{\mathrm{IR}}$ as reference.

{\it Herschel Space Observatory}\footnote{{\it Herschel} is an ESA space observatory with science instruments provided by European-led Principal Investigator consortia and with important participation from NASA.} (\citealt{Pilbratt2010b}) provides observations with excellent angular resolution for many famous and well-studied extragalactic fileds from far-infrared (FIR) to submillimeter band. Photodetector Array Camera and Spectrometer (PACS; \citealt{Poglitsch2010}) onboard {\it Herschel} observed the FIR sky in 70, 100 and 160\,$\mathrm{\mu m}$ bands, while Spectral and Photometric Imaging Receiver (SPIRE; \citealt{Griffin2010}) present maps of the submillimeter sky in 250, 350 and 500\,$\mathrm{\mu m}$ bands. Combining with near-infrared (NIR) and mid-infrared (MIR) observations from {\it Spitzer Space Telescope} and optical data from other telescopes, such as {\it Hubble Space Telescope} ({\it HST}), we are able to perform a multi-wavelength SED fitting from UV to submillimeter. Because dust emission results from dust attenuation in optical band, we believe that SED fitting with constraints from optical observations could provide a more actual estimate of $L_{\mathrm{IR}}$ relative to that of fitting only included IR constraints as which is done by most of previous works (e.g., \citealt{Elbaz2010,Galametz2013}).

In this paper, we use photometric data of the Great Observatories Origins Deep Survey northern (GOODS-North) and southern (GOODS-South) fields from UV to submillimeter to study the template-based monochromatic extrapolated total infrared luminosities $L_{\mathrm{IR}}$. Here, $L_{\mathrm{IR}}$ is defined as integration of the SED over the 8-1000\,$\mathrm{\mu m}$ wavelength range. We first compare the differences between the extrapolated $L_{\mathrm{IR}}$ from the bands that were frequently used (\citealt{Elbaz2010,Elbaz2011,Wuyts2011a}) based on different templates. A multi-wavelength SED fitting is taken in order to obtain a reference $L_{\mathrm{IR}}$ which is then used to compare with the extrapolated values from different templates. We also repeat this comparison in the rest-frame to study how well templates can describe the IR emission of galaxies. \par
This paper is organized as follows. Section 2 presents a brief description of our multi-wavelength data. In Section 3, we introduce our method to calculate the monochromatic extrapolated $L_{\mathrm{IR}}$ and the code used for SED fitting. We then analyse our results and put forward some main conclusions in Section 4. We discuss how factors such as confusion noise of observations influence our conclusions in Section 5 and give a short summary in Section 6. Throughout this paper, we adopt a \cite{Chabrier2003} initial mass function (IMF) and a $\Lambda$CDM cosmology with $H_0=70$\,km\ s$^{-1}$\,Mpc$^{-1}$, $\Omega_{\Lambda}=0.7$, $\Omega_{\mathrm{m}}=0.3$.

\section{Multi-wavelength Data}
\label{sect:data}

Benefiting from the wealth of multi-wavelength data of the GOODS-North and GOODS-South fields, we are able to compile an UV-to-submillimter catalog for {\it Herschel} detected sources. Our multi-wavelength catalog is mainly comprised of observations from three surveys: the 3D-HST survey (\citealt{Brammer2012}), the PACS Evolutionary Probe (PEP; \citealt{Lutz2011}) survey and the {\it Herschel} Multi-tiered Extragalactic Survey (HerMES; \citealt{Oliver2012}).

\subsection{Optical and NIR data}
Optical and NIR data is extracted from the 3D-HST catalogs of the v4.1 data release (\citealt{Brammer2012,Skelton2014}), which is publicly released at the survey's website\footnote{http://3dhst.research.yale.edu/Home.html}. 3D-HST is a NIR spectroscopic survey with {\it HST}, encompasses the same five well-studied extragalactic fields of the Cosmic Assembly Near-infrared Deep Extragalactic Legacy Survey (CANDELS; \citealt{Grogin2011,Koekemoer2011}) that provided photometric observations. Combining with a vast array of ancillary publicly available photometric data of the CANDELS/3D-HST fields from other surveys, \cite{Skelton2014} performed a photometric measurement to create multi-wavelength catalogs that cover a wavelength range of 0.3--8\,$\mathrm{\mu m}$. For the fields we concerned, except for the observations from {\it HST}/WFC3 ({\it F{\rm 125}W, F{\rm 140}W, F{\rm 160}W}), \cite{Skelton2014} used additional images in 19 bands for GOODS-North catalog and 37 bands for GOODS-South catalog. Note that we do not include all these bands in our study, see Table \ref{tab1} and the related text for details.

The 3D-HST catalogs contain the total fluxes for all available bands and the aperture fluxes in 0.7 arcsec for the {\it F{\rm 140}W} and {\it F{\rm 160}W} filters. Nevertheless, we only make use of the total fluxes in this study. Both spectroscopic redshifts (if available) and photometric redshifts, which were determined by the {\sc EAZY} code\footnote{https://github.com/gbrammer/eazy-photoz/} (\citealt{Brammer2008c}), are also provided in the catalogs. With the aim of studying the infrared luminosities of galaxies, we only consider sources with a flag of $\mathtt{use\_phot}=1$ in the catalog, which labels galaxies that have reasonably uniform quality of photometry as well as well-derived physical properties and photometric redshifts (\citealt{Skelton2014}).

\subsection{MIR and FIR data}
Our MIR and FIR data is from the first public data release (DR1) of the PEP survey\footnote{http://www.mpe.mpg.de/ir/Research/PEP/DR1}(\citealt{Lutz2011,Magnelli2013}). \cite{Magnelli2013} presented the deepest FIR images and catalogs of the GOODS-North and GOODS-South fields, which were constructed by using combined {\it Herschel}/PACS data from the PEP survey and GOODS-Herschel survey (\citealt{Elbaz2011}). The DR1 of PEP survey provided two kinds of catalogs for each field, one is constructed using the positions of the {\it Spitzer}/MIPS 24\,$\mathrm{\mu m}$ detected sources as priors to extract sources in the PACS maps, the other is a ``blind'' catalog that created by point spread function (PSF) fitting without any positional priors. Here we only make use of the prior source catalogs, which contain 24\,$\mathrm{\mu m}$ flux densities, PACS measurements down to 3$\sigma$ significance.

These catalogs also provide a ``clean index'', which is a measurement of the number of bright neighbours  around a given sources in MIPS 24\,$\mathrm{\mu m}$, PACS 100\,$\mathrm{\mu m}$ and 160\,$\mathrm{\mu m}$ bands. Here, we label sources with $\mathtt{clean\_index}\leq 1$ as clean sources for sources that only one or no bright source closer than 20 arcsec in 24\,$\mathrm{\mu m}$ map and no bright source closer than 6.7, 11 arcsec in 100, 160\,$\mathrm{\mu m}$ maps, respectively. Bright sources are defined as sources whose flux densities are brighter than half of that of the given source. Sources with $\mathtt{clean\_index}> 1$ are labelled as non-clean sources which means that these sources may suffer contamination from their bright neighbours (\citealt{Magnelli2013}). It should be noted that the PACS observations of the GOODS-South field included 70, 100 and $160\,\mathrm{\mu m}$ bands, but observation from $70\,\mathrm{\mu m}$ is unavailable for the GOODS-North field. As a result, the following study that limited to $70\,\mathrm{\mu m}$ band only make use of data from GOODS-South field.

\subsection{Submillimeter data}
Our submillimeter data is limited to the {\it Herschel}/SPIRE observations, i.e., 250, 350 and 500\,$\mathrm{\mu m}$ bands. We use the band-merged catalogs based on 250\,$\mathrm{\mu m}$ positions from the second Data Release (DR2) of the HerMES data\footnote{http://hedam.lam.fr/HerMES/} (\citealt{Wang2014}). Due to the update of the SPIRE calibration software after the DR2, flux correction of 1.0253, 1.0250 and 1.0125 at 250, 350 and 500\,$\mathrm{\mu m}$, respectively, are applied to the photometry in the released catalogs, as recommended by \cite{Wang2014}.

After a cross-matching between these catalogs, we compile a sample of 894 sources. Among these soures, 377 sources ($\sim42\%$) have  spectroscopic redshifts. Galaxies with AGN signature are also identified from the sample. \cite{Xue2011} provided a detailed {\it Chandra} source catalog for the 4\,Ms {\it Chandra} Deep Field-South (CDF-S), which contained 740 X-ray sources. Because the GOODS-South field is one part of the CDF-S, we cross-match our sample with X-ray sources classified as AGN in the catalog of \cite{Xue2011} to identify AGNs. For GOODS-North field, which is covered by the {\it Chandra} Deep Field North (CDF-N), AGNs are selected based on X-ray point-source catalog presented by \cite{Alexander2003} for the 2\,Ms CDF-N. In this field, an X-ray source with an intrinsic X-ray luminosity of $L_{0.5-8\,\mathrm{keV}}\geq 3\times10^{42}\,\mathrm{erg\,s^{-1}}$ or an effective photon index of $\Gamma\leq 1.0$ is classified as an AGN (\citealt{Xue2011}). After this selection, our sample is separated into three populations: 321 clean sources, 438 non-clean sources and 135 AGNs.

\section{Methods}
\label{sect:methods}

\subsection{Calculating the monochromatic extrapolated luminosity}
We compare three different IR SED templates: CE01, DH02 and W08. CE01 and DH02 templates are two of the most widely used IR SED templates. They are both luminosity-dependent and locally calibrated (i.e., calibrated by local galaxy sample). CE01 template was generated to reproduce the observed relations between different MIR and FIR luminosities based on observations of 0.44--100\,\um\ and 850\,\um\ of nearby galaxies. There are four original SEDs, which were created to describe the UV to submillimeter radiation of galaxies with four different luminosity classes (ultra-luminous infrared galaxies (ULIRGs), luminous infrared galaxies (LIRGs), starbursts, and normal galaxies) using the \citet{Silva1998} models. These SEDs were interpolated to generate a series of templates with intermediate luminosities, in which the SED best fits the predicted luminosities from the observed relations was selected as final template for each luminosity bin. These final templates are stored in an IDL save file and publicly available\footnote{http://david.elbaz3.free.fr/astro\_codes/chary\_elbaz.html}. DH02 template was built to fit the observed color-color relations between local normal galaxies and originally constructed by \citet{Dale2001}, which presented a wide range of semiempirical IR SEDs for different heating levels of interstellar environment. These SEDs are parameterized as $\mathrm{d}M_{\mathrm{d}}(U)\propto U^{-\alpha}\mathrm{d}U$, where $M_{\mathrm{d}}(U)$ represents the dust mass heated by an interstellar radiation field with an intensity of $U$ and the index $\alpha$, which ranges from 1 to 2.5 for normal galaxies, describes the relative contributions of radiation fields with different intensities. This template was calibrated by using observations of normal star-forming galaxies between 3 and 100\,$\mu$m in \citet{Dale2001}, and improved the $\lambda>100\,\mu$m part by observations from FIR and submilimeter bands in DH02. These SEDs are saved as a function of {\it IRAS} color and released online\footnote{http://physics.uwyo.edu/$\sim$ddale/research/seds/seds.html}.

CE01 template is consisted of 105 SEDs and each SED is attributed to a fixed $L_{\mathrm{IR}}$. We use the standard CE01 technique to calculate the monochromatic extrapolated luminosity for a given band. The main steps are as follows:
\begin{enumerate}
\renewcommand{\labelenumi}{(\theenumi)}
\item Shift all the SEDs in the template to the redshift of the galaxy, convolve them with the response function of the given filter to get a series of templated fluxes $F_{\nu,\mathrm{temp}}$.
\item Convolve the assumed SED shape that used in extracting flux density (e.g., a blackbody spectrum with a temperature of 10,000\,K for {\it Spitzer}/MIPS 24\,$\mathrm{\mu m}$ band\footnote{http://irsa.ipac.caltech.edu/data/SPITZER/docs/mips/mipsinstrumenthandbook/}) with the response function of the filter to get the reference flux $F_{\nu,\mathrm{ref}}$ and a series of scaled factors $\mathrm{fac}=F_{\nu,\mathrm{temp}}/F_{\nu,\mathrm{ref}}$.
\item Interpolate the assumed SED shape at the nominal wavelength $\lambda_{\mathrm{nom}}$ (e.g., $\lambda_{\mathrm{nom}}=23.675\,\mathrm{\mu m}$ for {\it Spitzer}/MIPS 24\,$\mathrm{\mu m}$ band) and multiply by the scaled factors to obtain the templated flux densities $f_{\nu,\mathrm{temp}}$ corresponding to all the SEDs.
\item Interpolate $f_{\nu,\mathrm{temp}}$ to find the SED that gives the observed flux density $f_{\nu,\mathrm{obs}}$, integrate the 8-1000$\,\mu$m range of this SED to obtain $L_{\mathrm{IR}}$.
\end{enumerate}
We apply the above method for each source and each MIR-to-submillimeter band in our sample to find the best SED and calculate the corresponding extrapolated $L_{\mathrm{IR}}$.

DH02 template includes 64 SEDs represented a wide range of dust heating intensities. We use the same technique to find the observed SED of DH02 template. Because SEDs of DH02 were saved in arbitrary unit but parameterized by the {\it IRAS} $f_{\nu}(60\,\mathrm{\mu m})/f_{\nu}(100\,\mathrm{\mu m})$ FIR color, we determine $L_{\mathrm{IR}}$ using the \cite{MECDGM2006} relation between the $f_{\nu}(60\,\mathrm{\mu m})/f_{\nu}(100\,\mathrm{\mu m})$ ratio and $L_{\mathrm{IR}}$:
\begin{equation}
\log_{10}[f_{\nu}(60\,\mathrm{\mu m})/f_{\nu}(100\,\mathrm{\mu m})]=0.128\times \log_{10}L_{\mathrm{IR}}-1.611.
\end{equation}
\cite{Dale2014} updated the DH02 template by using an average template of normal galaxies acquired from {\it Spitzer} to replace the mid-infrared part of the original template. Hence we adopt this updated version in our work.

W08 template is a single luminosity-independent SED which was constructed from DH02 template. As mentioned above, DH02 template includes 64 SEDs parameterized by the index $\alpha$. The logarithm of all these SEDs were averaged to derived W08 template \citep{Wuyts2011a}. In other words, W08 template is the logarithmic mean of SEDs of normal galaxies with different active levels. This template exhibits a good feature that it can result in a consistency between 24\,\um\ and PACS-derived $L_{\mathrm{IR}}$, while CE01 and DH02 both overestimate the PACS-derived $L_{\mathrm{IR}}$ at $z\gtrsim2$ \citep{Wuyts2011a,Wuyts2011b}. The template and a table contained conversion factors from MIPS 24\,\um\ and PACS 70\,\um, 100\,\um, and 160\,\um\ to $L_{\mathrm{IR}}$ is available online\footnote{http://www.mpe.mpg.de/$\sim$swuyts/Site/Lir\_template.html}. To obtain a W08 $L_{\mathrm{IR}}$, we follow the first three steps of CE01, as described above, to get a template flux density $f_{\nu,\mathrm{temp}}$, then scale the SED by a factor of $f_{\nu,\mathrm{obs}}/f_{\nu,\mathrm{temp}}$ and integrate it.

\subsection{SED fitting}
\label{subsect:fit}

To check the accuracy of these monochromatic extrapolated luminosities, we carry out a multi-wavelength SED fitting using the {\sc MAGPHYS} code\footnote{http://www.iap.fr/magphys/magphys/MAGPHYS.html} (\citealt{DaCunha2008}). {\sc MAGPHYS} allows a fitting covering a wide wavelength range from UV to submillimeter, employs an energy balance technique to consistently connect the dust emission with the attenuation of stellar emission. It can account for various star formation histories (SFHs) as well as complex dust configuration. The star formation history is described by a continuous model superimposed by random star formation bursts. The underlying continuous component is paramterized as an exponentially declining form with star formation rate $\psi(t)\propto e^{-\gamma t}$, where $t$ is the time since the onset of star formation and $\gamma$ is the star formation timescale parameter in Gyr$^{-1}$. Dust emission assumed by the code is consisted of different components including a fixed template spectrum for the polycyclic aromatic hydrocarbons (PAH) emission, emission from stochastically heated small grains, emission from warm and cold grains in thermal equilibrium. The reliability of {\sc MAGPHYS} was investigated by \cite{Hayward2015}, which found that {\sc MAGPHYS} can recover most physical parameters of the simulated galaxies well when the true attenuation curve is relatively consistent with that assumed by the code. This result indicates that {\sc MAGPHYS} can effectively extract information of the observed sources from multi-wavelength data, which enables us to investigate the accuracy of the monochromatic extrapolated $L_{\mathrm{IR}}$.

{\sc MAGPHYS} assumes that dust is heated only by stellar emission. As a result, any possible contribution from AGN component would be ignored in our fitting. However, \cite{Hayward2015} found that the galaxy properties derived from {\sc MAGPHYS} are generally unaffected by AGN component, even if the AGN contribution is as much as 25 percent of the UV-to-millimeter luminosity. As recommended by the code's release website, we apply stellar population synthesis model of \cite{Bruzual2003}, hereafter BC03, to calculate stellar emission instead of its CB07 version (unpublished), which is the default templates used in {\sc MAGPHYS}.
Due to the incompleteness of the built-in filters of {\sc MAGPHYS}, we could not make use of all the available photometric data in SED fitting. For galaxies in the GOODS-North field, we include 26 bands in fitting at most, while the same number is 41 for galaxies in the GOODS-South field, as is shown in Table \ref{tab1}.
\placetable{tab1}

\section{Results}
\label{sect:results}

\subsection{Comparison between IR SED templates}
\begin{figure}
\centering
\plotone{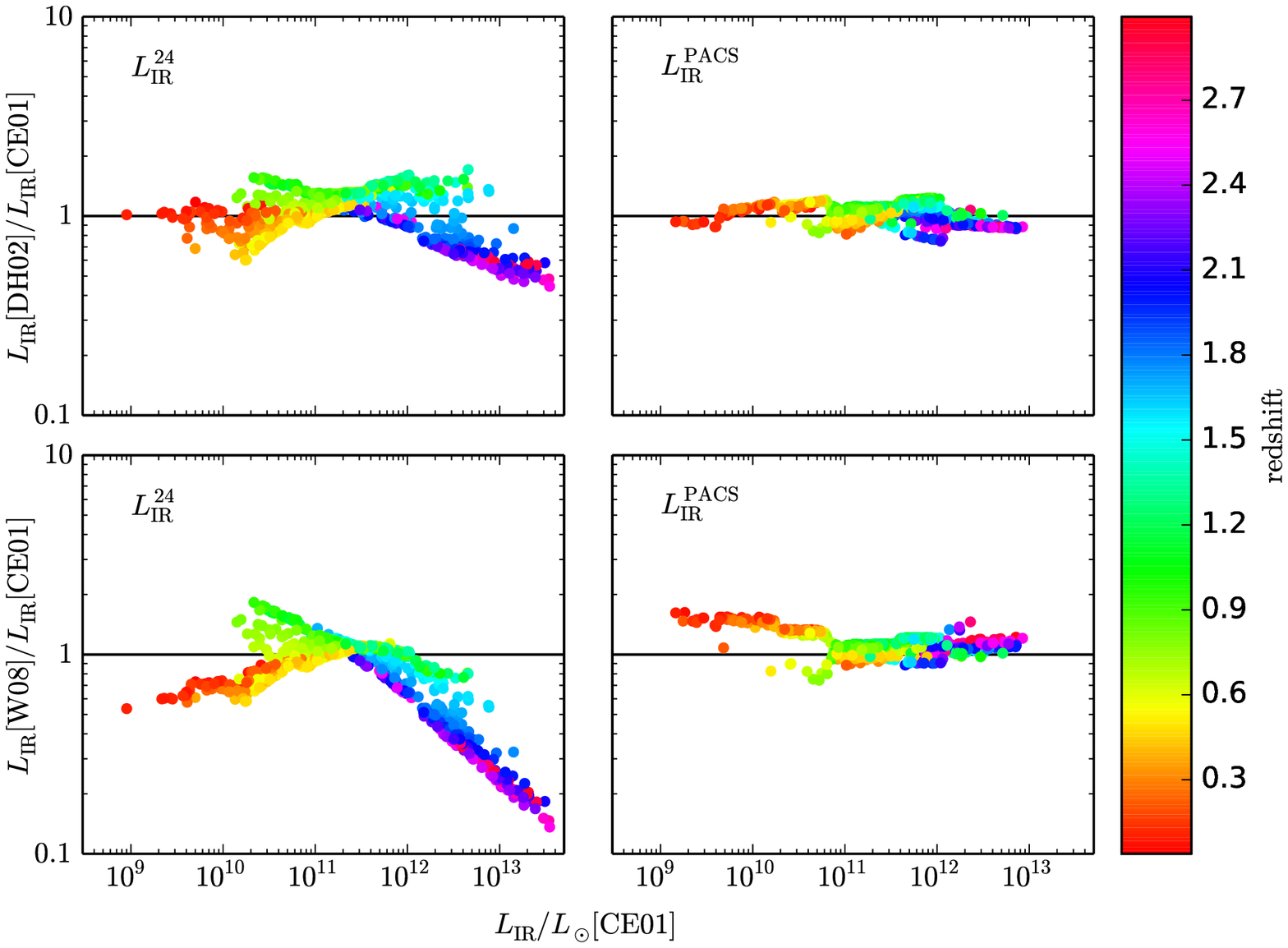}
\caption{Comparison of monochromatic extrapolated luminosity from {\it Spitzer}/MIPS 24\,$\mathrm{\mu m}$ $L_{\mathrm{IR}}^{24}$ and {\it Herschel}/PACS bands $L_{\mathrm{IR}}^{\mathrm{PACS}}$ based on CE01, DH02 and W08 templates. The color is encoded by the redshifts of galaxies.}
\label{fig1}
\end{figure}

We compare the difference between the monochromatic extrapolated $L_{\mathrm{IR}}$ derived from {\it Spitzer}/MIPS 24\,$\mathrm{\mu m}$ ($L_{\mathrm{IR}}^{24}$) and {\it Herschel}/PACS band ($L_{\mathrm{IR}}^{\mathrm{PACS}}$) based on different IR SED templates. Here, $L_{\mathrm{IR}}^{\mathrm{PACS}}$ is defined as extrapolation from the available longest wavelength PACS band that has a significant ($>3\sigma$) detection (\citealt{Wuyts2011a}). This comparison is shown in Figure \ref{fig1}, and the color indicates the redshifts of galaxies.
\placefigure{fig1}

For $L_{\mathrm{IR}}^{24}$, the result of DH02 template is similar to that of CE01 template, but slightly lower for galaxies at $z\gtrsim 2$, which indicates a different emission intensity at $\lambda\sim 8\,\mathrm{\mu m}$ between these two templates. The result of W08 shows more differences with that of CE01 or DH02, for which the low-luminosity end and high-luminosity end are both notably systematic lower. In the case of $L_{\mathrm{IR}}^{\mathrm{PACS}}$, both DH02 and W08 template give nearly the same result as that of CE01, but $L_{\mathrm{IR}}^{\mathrm{PACS}}$ from W08 is slightly higher than that of CE01 when $L_{\mathrm{IR}}<10^{11}L_{\odot}$ or $z\lesssim 0.5$.
Therefore, CE01 and DH02 templates present nearly the same extrapolation from MIR to FIR, except for a weaker emission at $\lambda\sim 8\,\mathrm{\mu m}$ for DH02 template, while W08 template exhibits more differences with the above two. In consideration of the accuracy of template-based luminosity of CE01 which is discussed below, we suggest that the monochromatic extrapolated $L_{\mathrm{IR}}$ derived from W08 template should be used with caution, especially for galaxies with redshift $z\lesssim 0.5$, although the differences are quite small.

\subsection{SED fitting result}

\begin{figure}
\centering
\plotone{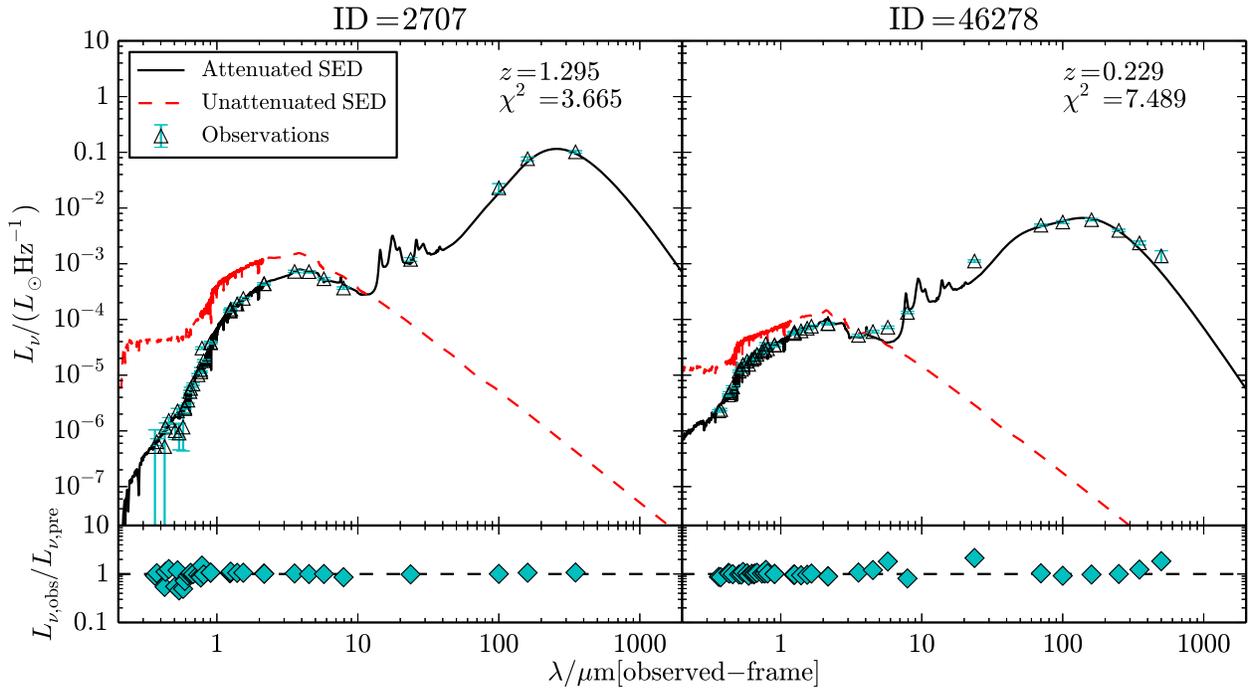}
\caption{Example of {\sc MAGPHYS} fitting results for two randomly selected galaxies. The upper panels show the comparisons between the best-fit SED models and the observed data. The black solid lines and the red dashed lines represent the attenuated and unattenuated best-fit SED models, respectively. The triangles with error bars show the measured monochromatic luminosities and their 1$\sigma$ errors. The redshift of the sources and the $\chi^2$ of the best-fit models are shown at the upper right corner. The bottom panels present the ratio of the observed luminosities to the predicted values of the best-fit SED models as a function of the observed-frame wavelength.}
\label{fig2}
\end{figure}

{\sc MAGPHYS} performs acceptable fits for most galaxies in our sample. For example, in Figure \ref{fig2}, we plot the best-fit models (i.e., the models given the minimal $\chi^2$) for two galaxies selected randomly from our sample. Their best-fit models have small $\chi^2$, while the SEDs show a good consistency with the observations in all available bands.
\placefigure{fig2}

\begin{figure}
\centering
\includegraphics[scale=0.8]{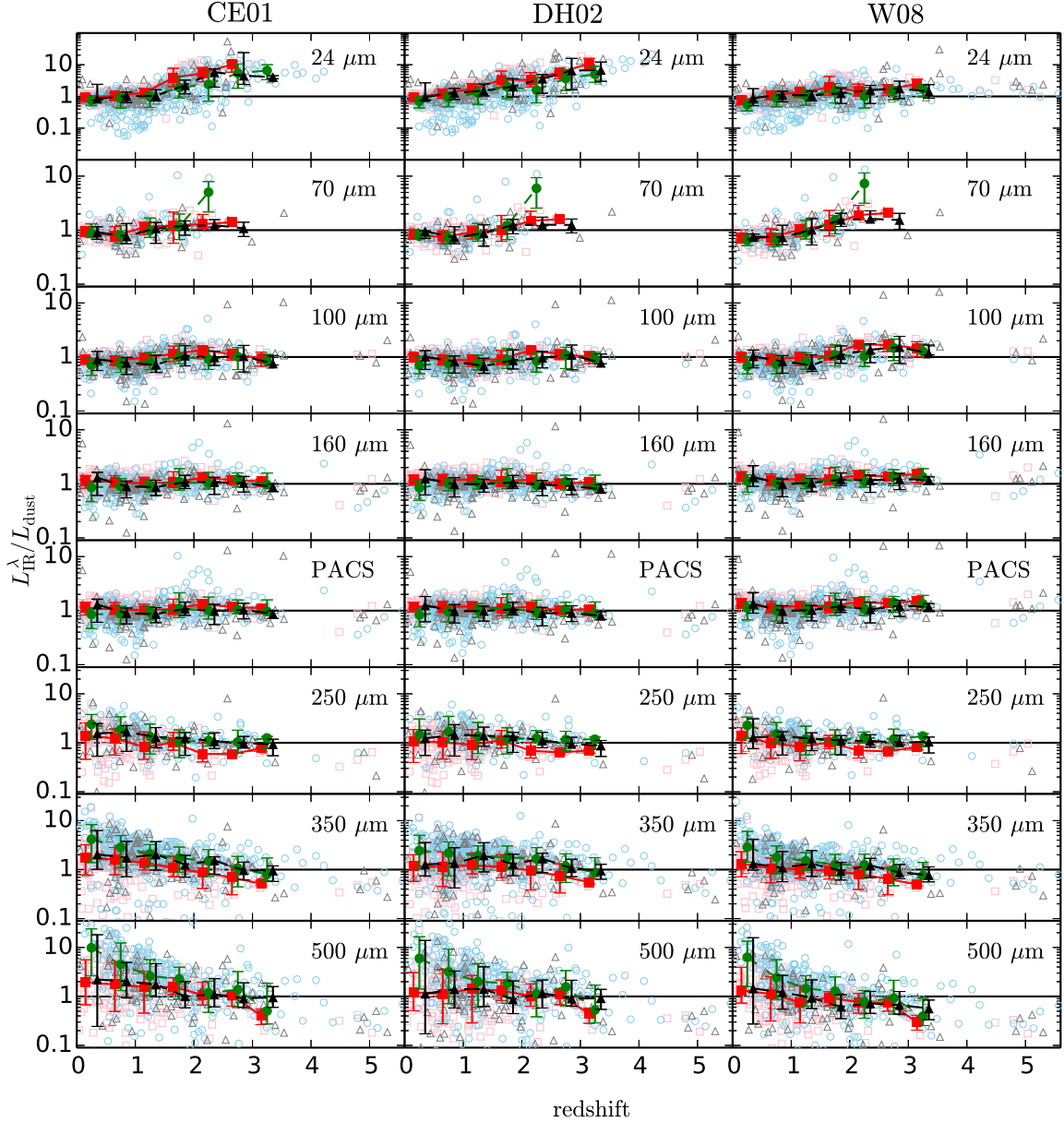}
\caption{The ratio of $L_{\mathrm{IR}}$ derived from monochromatic measurements to {\sc MAGPHYS} best-fit result $L_{\mathrm{dust}}$ as a function of redhsift for CE01 template ({\it left panel}), DH02 template ({\it middle panel}) and W08 template ({\it right panel}). The squares, circles and triangles represent clean sources, non-clean sources and AGNs, respectively. Symbols with lines show the binned median values and error bars represent the 16th and 84th percentiles of the distributions. Note that the bins of these three subsamples are the same, but the median curves are shifted in order to show clearly.}
\label{fig3}
\end{figure}

The comparison between the template-based monochromatic extrapolated IR luminosity, denoted as $L_{\mathrm{IR}}^{\lambda}$ ($\lambda=$24, 70, 100, 160, 250, 350, 500\,$\mathrm{\mu m}$), and the output of the best-fit model $L_{\mathrm{dust}}$ is as Figure \ref{fig3}.  $L_{\mathrm{IR}}^{\mathrm{PACS}}$ are also plotted for comparing. The squares, circles and triangles show the distributions of  clean sources (only one or no close bright neighbors), non-clean sources (more than one close bright neighbors) and AGNs, respectively. For sources with redshifts $z<3.5$, we divide them into seven redshift bins, calculate the median value of $L_{\mathrm{IR}}^{\lambda}/L_{\mathrm{dust}}$ and the 16th to 84th percentile range for each bin. The symbols with error bars and lines in Figure \ref{fig3} represent these binned parameters. Note that the bins of these three populations are the same, although the median curves are shifted so as to show clearly. In Table \ref{tab2}, we separate our clean sources into two subsample according to their redshifts, and provide a quantitative description of the result in Figure \ref{fig3}.

\placefigure{fig3}

For $L_{\mathrm{IR}}^{\lambda}$ derived from 24--160\,$\mathrm{\mu m}$, the $L_{\mathrm{IR}}^{\lambda}/L_{\mathrm{dust}}$ ratio of clean sources, non-clean sources and AGNs show a similar redshift evolution out to $z\sim 3.5$, regardless of template. It implies that the close bright neighbors of these non-clean sources have almost no contamination on them in 24--160\,$\mathrm{\mu m}$ bands. As for AGNs, this consistency indicates that the IR luminosities of most AGNs might be dominated by star formation activities of their host galaxies, which agrees with previous studies (\citealt{Elbaz2010,Elbaz2011}) for sources with $z\lesssim 2.5$.

In the case of CE01 template, $L_{\mathrm{IR}}^{24}$ exhibits a good estimate of $L_{\mathrm{dust}}$ at $z<1.5$ with a dispersion of about $35\%$ for clean sources, but shows a significant overestimate by a factor of about 3--8 when $z\geq 1.5$. This discrepancy between $L_{\mathrm{IR}}^{\lambda}$ and $L_{\mathrm{dust}}$ has been known as ``mid-IR excess'' by previous works (\citealt{Daddi2007,Papovich2007,Elbaz2010}).  One possible explaination is that using IR templates calibrated by compact starbursts to estimate $L_{\mathrm{IR}}$ of galaxies with extended star formation is the origin of this excess problem (\citealt{Elbaz2011}).
Furthermore, extrapolations from FIR bands (70--160\,$\mathrm{\mu m}$) provide a better estimate of $L_{\mathrm{dust}}$ with a small dispersion when $z<3.5$, especially for clean sources. The median values and the 16th--84th percentile ranges of $L_{\mathrm{IR}}^{70}/L_{\mathrm{dust}}$, $L_{\mathrm{IR}}^{100}/L_{\mathrm{dust}}$, $L_{\mathrm{IR}}^{160}/L_{\mathrm{dust}}$, and $L_{\mathrm{IR}}^{\mathrm{PACS}}/L_{\mathrm{dust}}$ for all clean sources are $0.94_{-0.27}^{+0.37}$, $0.92_{-0.21}^{+0.25}$, $1.11_{-0.26}^{+0.31}$, and $1.10_{-0.25}^{+0.33}$, respectively. \cite{Elbaz2010} presented a similar result applying a different reference $L_{\mathrm{IR}}$, which was determined from the best-fit of the normalized infrared SED templates (e.g., CE01, DH02) using at least two photometric measurements above rest-frame 30\,$\mathrm{\mu m}$. We stress that their result demonstrated the self-consistency of templates in FIR bands, while ours prove the accuracy of this extrapolation method in the same bands. Several sources at high redshift imply that the tight correlation between $L_{\mathrm{IR}}^{\lambda}$ derived from PACS bands and $L_{\mathrm{dust}}$ might be held even at $z\sim5$. On the other hand, the flat curves of median values out to $z\sim 3.5$ show that no evidence indicate an evolution in this redshift interval.
However, from 250\,$\mathrm{\mu m}$ to 500\,$\mathrm{\mu m}$, $L_{\mathrm{IR}}^{\lambda}$ tend to overestimate $L_{\mathrm{dust}}$ at low redshift and underestimate $L_{\mathrm{dust}}$ at high redshift. The non-clean sources always have a larger $L_{\mathrm{IR}}^{\lambda}/L_{\mathrm{dust}}$ ratio relative to that of  clean sources, which is due to the serious confusion noise of SPIRE bands. Although \cite{Wang2014} carefully extracted point sources from observations of all HerMES fields, the non-clean sources still suffer serious contamination by their bright neighbors or the background noise. As a result, none of $L_{\mathrm{IR}}^{\lambda}$ from these submillimeter bands can provide an estimate that is as good as PACS bands.

The $L_{\mathrm{IR}}^{\lambda}/L_{\mathrm{dust}}$ ratios from 24--160\,$\mathrm{\mu m}$ based on DH02 template are similar to that of CE01 template, with the median values and $68\%$ dispersions of  $0.82_{-0.19}^{+0.36}$, $0.94_{-0.23}^{+0.20}$, $1.19_{-0.26}^{+0.28}$, $1.18_{-0.27}^{+0.28}$ for all clean sources in 70, 100, 160\,$\mathrm{\mu m}$ bands and PACS, respectively. In both Figure \ref{fig3} and Table \ref{tab2}, a slight difference of the $L_{\mathrm{IR}}^{70}/L_{\mathrm{dust}}$ ratio can be found for clean sources between those with redshifts of samller and larger than 1.5. With respect to SPIRE bands, the non-clean sources still provide a larger $L_{\mathrm{IR}}$ relative to $L_{\mathrm{dust}}$ because of the confusion noise. Nevertheless, we also note that for clean sources with $z<2$, $L_{\mathrm{IR}}^{\lambda}$ from these submillimeter bands can give a nearly unbiased estimation of $L_{\mathrm{dust}}$, although the dispersion is large. Their median values and the corresponding dispersions are $1.02_{-0.56}^{+0.82}$, $1.19_{-0.61}^{+0.92}$, $1.22_{-0.88}^{+1.71}$, respectively. Therefore, DH02 template performs better than CE01 template in these submillimeter bands.

With regard to W08 template, $L_{\mathrm{IR}}^{24}$ doesn't suffer a significant mid-IR excess problem, which is consistent with \cite{Wuyts2011a}, but still shows a weak trend that the $L_{\mathrm{IR}}^{24}/L_{\mathrm{dust}}$ ratio increases as redshift increases. Table \ref{tab2} also shows that the median value of this ratio of clean sources with $z\geq 1.5$ is 88\% larger than that of clean sources with $z<1.5$. However, all extrapolations from PACS bands could not provide an estimate as good as that of CE01 or DH02 template. Concretely, the evolution trend of the $L_{\mathrm{IR}}^{70}/L_{\mathrm{dust}}$ ratio of W08 is similar to that of DH02 but shows more significant deviation from unity in both low and high redshift regions. For clean sources with $z\geq 1.5$, extrapolations from 100\,\um\ band tend to overestimate $L_{\mathrm{dust}}$ by 30\%. Furthermore, $L_{\mathrm{IR}}^{160}$ also shows a systematic overestimate of $27\%$ comparing with $L_{\mathrm{dust}}$. In submillimeter bands, extrapolations based on W08 template perform as well as DH02, i.e., $L_{\mathrm{IR}}^{\lambda}$ can provide a rough estimate of $L_{\mathrm{dust}}$ with a large dispersion for clean sources with $z\lesssim 2$.

In conclusion, among the three templates, CE01 template provides the best estimate of $L_{\mathrm{dust}}$ in PACS bands, while DH02 and W08 templates perform better in SPIRE bands, and only W08 template does not suffer a serious mid-IR excess problem. For the purpose of estimate $L_{\mathrm{IR}}$, we suggest that $L_{\mathrm{IR}}^{\mathrm{PACS}}$ proposed by \cite{Wuyts2011a} based on CE01 template can be a good estimator of $L_{\mathrm{dust}}$. Besides, if PACS (e.g., FIR bands) measurement is unavailable, extrapolations from SPIRE observations based on DH02 or W08 template can also give a rough estimate of $L_{\mathrm{dust}}$ for clean sources at $z\lesssim 2$. From the above conclusions, we can see that it is possible to construct a luminosity-dependent IR SED template, which can provide unbiased luminosity estimate in all MIR-to-submillimeter bands.\par

\subsection{How well do the templates describe the infrared emission of galaxies?}
\begin{figure}
\centering
\includegraphics[scale=0.8]{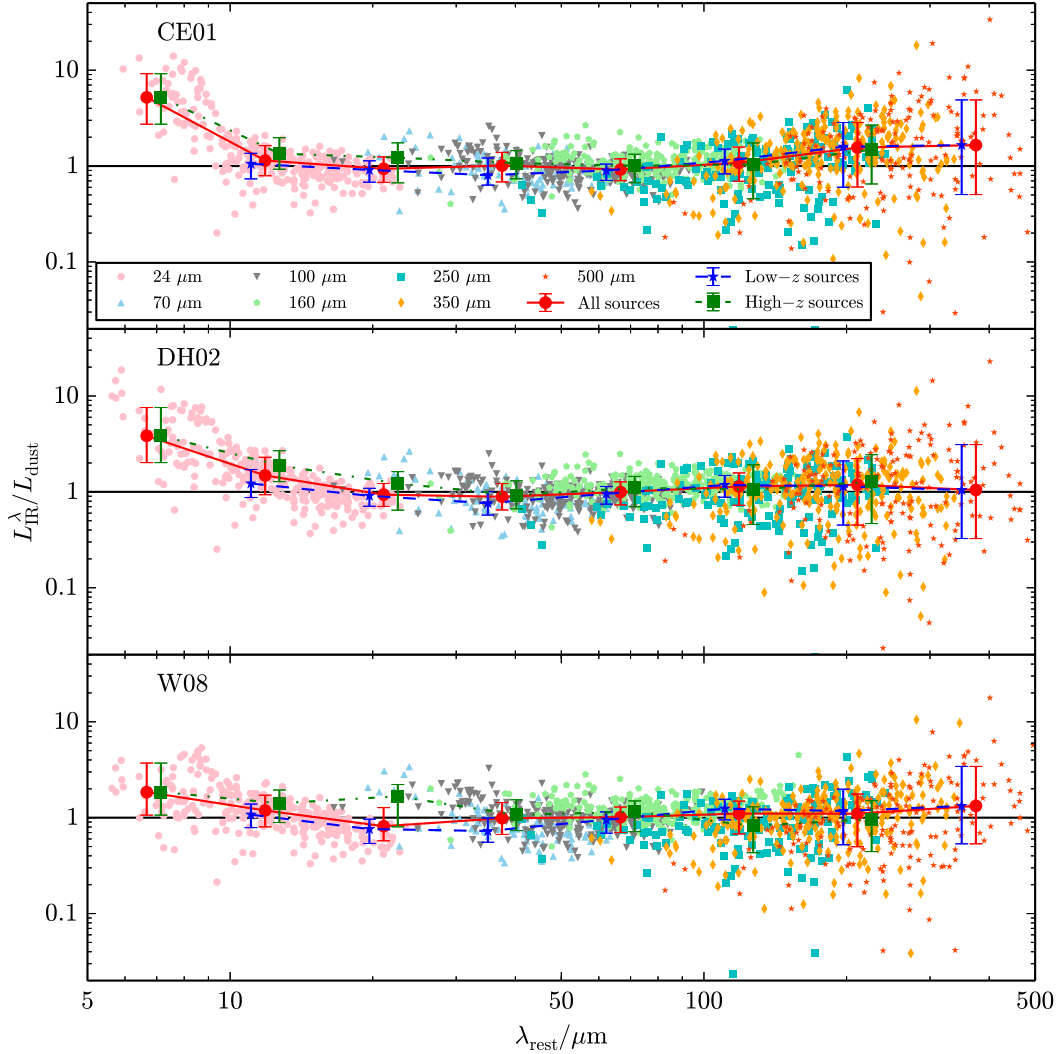}
\caption{The $L_{\mathrm{IR}}^{\lambda}/L_{\mathrm{dust}}$ ratio versus the rest-frame wavelength $\lambda_{\mathrm{rest}}$ corresponding to the observed wavelength from which $L_{\mathrm{IR}}^{\lambda}$ derived for clean sources. From top to bottom are results from CE01, DH02 and W08 templates, respectively. Points show the distribution of $L_{\mathrm{IR}}^{\lambda}/L_{\mathrm{dust}}$ ratio from different observed bands: 24, 70, 100, 160, 250, 350, 500\,$\mathrm{\mu m}$. The circles connected by solid lines represent the binned median values for all clean sources and error bars show the 16th and 84th percentiles of the binned distributions. The stars and squares with lines are the same as the circles but for the low-redshift sources ($z<1.5$) and high-redshift sources ($z\geq 1.5$), respectively. The bins used are the same regardless of templates and samples, but the median curves are shifted in order to show clearly.}
\label{fig4}
\end{figure}

As mentioned above, extrapolations based on CE01 or DH02 template from 24\,$\mathrm{\mu m}$ measurements for galaxies at $z\gtrsim 1.5$ would result in an overestimate of actual infrared luminosity, while extrapolated $L_{\mathrm{IR}}$ from 70--160\,$\mathrm{\mu m}$ bands provide acceptable estimate out to $z\sim 3.5$. Hence it is necessary to determine the range of application for each SED template. The wide redshift range of our sample enable us to fully sample the whole 5--500\,$\mathrm{\mu m}$ rest-frame wavelength range. For all clean sources, we plot the relations between the $L_{\mathrm{IR}}^{\lambda}/L_{\mathrm{dust}}$ ratio and the rest-frame wavelength $\lambda_{\mathrm{rest}}$, which is corresponding to the observed band from which $L_{\mathrm{IR}}^{\lambda}$ is derived, in Figure \ref{fig4}. To find any evidence of an evolution effect, we also divide clean sources into two subsamples according to their redshifts: the low-redshift sources with $z<1.5$ and the high-redshift sources with $z\geq 1.5$. We separate the whole 5--500\,$\mathrm{\mu m}$ wavelength range into 8 bins and compute the median and the 16th--84th percentile range of each binned distribution for each population. This binned result is overplotted in Figure \ref{fig4}, and the corresponding values for all clear sources are shown in Table \ref{tab3}.
\placefigure{fig4}

Monochromatic extrapolated $L_{\mathrm{IR}}^{\lambda}$ based on CE01 template can provide an acceptable estimate of the actual infrared luminosity when $10\,\mathrm{\mu m}\lesssim \lambda_{\mathrm{rest}} \lesssim 100\,\mathrm{\mu m}$. For all extrapolations with $10\,\mathrm{\mu m}\leq \lambda_{\mathrm{rest}}\leq 100\,\mathrm{\mu m}$, the median value of $L_{\mathrm{IR}}^{\lambda}/L_{\mathrm{dust}}$ is 0.98 with a dispersion of about $30\%$. However, extrapolations from a rest-frame wavelength that shorter than 10\,$\mathrm{\mu m}$ would significantly overestimate the total infrared luminosity. There is a tendency that as the rest-frame wavelength decreases, the overestimate becomes more serious. This indicates an excess emission of the template SEDs at this wavelength range. Although the construction of the CE01 template included observations from this wavelength range (see CE01), \cite{Elbaz2011} argued that these observations come from local ultra-luminous infrared galaxies (ULIRGs) which exhibit a higher $L_{\mathrm{IR}}$ to rest-frame $L_{8}$ ($=\nu L_{\nu}[8\mathrm{\mu m}]$) ratio but remain a minority at both low redshift and high redshift. Therefore extrapolations from $\lambda_{\mathrm{rest}}\sim 8\,\mathrm{\mu m}$ result in an overestimate of the actual IR luminosities. Furthermore, observation with a corresponding $\lambda_{\mathrm{rest}}$ longer than 100\,$\mathrm{\mu m}$ would also give a statistically higher IR luminosity with a large dispersion. This large dispersion covers the acceptable range of the $L_{\mathrm{IR}}^{\lambda}/L_{\mathrm{dust}}$ ratio, but the risk of a wrong estimate increases as the dispersion gets larger. This $L_{\mathrm{IR}}^{\lambda}$ excess in submillimeter bands may be due to the lack of constraint from these bands (out to 500\,$\mathrm{\mu m}$) when constructing the template.

In the case of DH02 template, an acceptable $L_{\mathrm{IR}}$ can be given in a larger rest-frame wavelength range, i.e. $10\,\mathrm{\mu m}\lesssim \lambda_{\mathrm{rest}}\lesssim 500\,\mathrm{\mu m}$. For all extrapolations with $10\,\mathrm{\mu m}\leq \lambda_{\mathrm{rest}}<100\,\mathrm{\mu m}$, $100\,\mathrm{\mu m}\leq \lambda_{\mathrm{rest}}< 500\,\mathrm{\mu m}$, the median values and the 16th--84th percentile ranges are $1.01_{-0.30}^{+0.44}$, $1.14_{-0.61}^{+0.88}$, respectively. Thus, as mentioned above, DH02 template performs better than CE01 template in submillimeter bands in both observed-frame and rest-frame. This feature benefits from the improvement of the $\lambda>100$\,\um\ region in DH02. To constraint the emission of this wavelength range, CE01 only used the predictions from the observed 850\,\um\ monochromatic luminosity $\nu L_{\nu}$ (850\,\um)-FIR luminosity $L$(40--500\,\um) relation which exhibit a large dispersion, and fit the whole 20--1000\,\um\ range simultaneously with silmilar preditions from $\lambda\leq 100$\,\um\ part. However, DH02 modified their FIR region independently of the $\lambda<100$\,\um\ part using direct observations of 850\,\um\ as well as other FIR bands longer than 100\,\um\ from {\it ISO}. Obviously, DH02 set stricter constraint on the $\lambda>100$\,\um\ range than CE01, and results in a better description of the dust emission in this wavelength range. On the other hand, since W08 template is the logarithmic mean of SEDs in DH02,  it inherits this feature from DH02 template and present acceptable performance in submillimeter bands.

The apparent  agreement of $L_{\mathrm{IR}}^{24}$ based on W08 template with $L_{\mathrm{dust}}$ presented in Figure \ref{fig3} could not indicate an unbiased estimate of $L_{\mathrm{dust}}$. Moreover, the distribution in Figure \ref{fig4} present a tendency that the $L_{\mathrm{IR}}^{24}/L_{\mathrm{dust}}$ ratio slightly decreases as the corresponding $\lambda_{\mathrm{rest}}$ increases. As a result, $L_{\mathrm{IR}}^{24}$ slightly overestimate $L_{\mathrm{dust}}$ at high-redshift end and underestimate it for sources at $z\sim 0$. This suggests that the mid-IR excess problem still exist for W08 template, but not as serious as the other two. For this simple template, extrapolations from a moderate wavelength range of $30\,\mathrm{\mu m}\lesssim \lambda_{\mathrm{rest}}\lesssim 200\,\mathrm{\mu m}$ can be nearly consistent with $L_{\mathrm{dust}}$, while result of the longer wavelength slightly overestimate.

From the above discussion, we can summarize that all the three templates present different degrees of enhanced emission at $\lambda_{\mathrm{rest}}\lesssim 10\,\mathrm{\mu m}$ compared with most of galaxies in our sample, while the emission of 10--100\,$\mathrm{\mu m}$ range can be well described regardless of template. Only DH02 template shows a nearly unbiased estimate of the emission of the rest-frame submillimeter part. Moreover, synthesizing the comparisons between high-redshift and low-redshift sources of all the three templates, we find that  the high-redshift population exhibit a slightly higher $L_{\mathrm{IR}}^{\lambda}/L_{\mathrm{dust}}$ ratio at $\lambda_{\mathrm{rest}}\lesssim 30\,\mathrm{\mu m}$, which hints a potential stronger emission for our high-redshift population. However, no evidence implies a redshift evolution of the rest-frame MIR-to-submillimeter SED of galaxies.

\section{Discussion}

\begin{figure}
\centering
\plotone{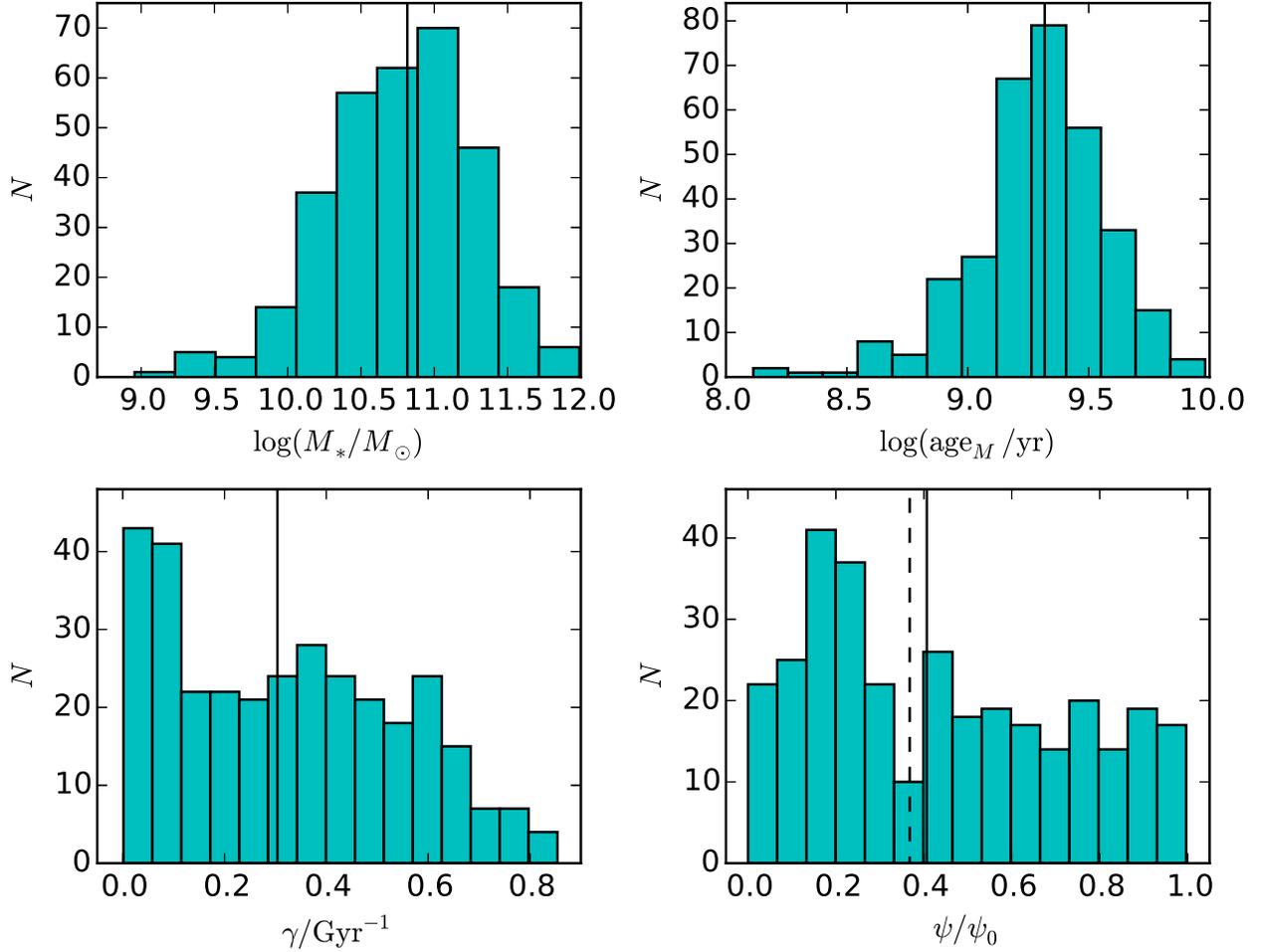}
\caption{Distribution of the stellar mass $M_*$, mass-weighted age $\mathrm{age}_M$, star formation timescale parameter $\gamma$, and the present star formation rate to initial star formation rate ratio $\psi/\psi_0$ derived from SED fitting employing BC03 stellar population models for our clean sample. The solid lines indicate the median values for each distribution, while the dashed line in the $\psi/\psi_0$ distibution shows the value of $\mathrm{e}^{-1}\approx 0.37$.}
\label{fig5}
\end{figure}

To check the self-consistency of the {\sc MAGPHYS} results, we plot the distributions of stellar mass $M_*$, mass-weighted age $\mathrm{age}_M$, star formation timescale parameter $\gamma$, and the present star formation rate to initial star formation rate ratio $\psi/\psi_0$ derived from SED fitting for our clean sample in Figure \ref{fig5}. The median values of each distribution are labelled by vertical solid lines, and the dashed line in the $\psi/\psi_0$ distibution marks the value of $\mathrm{e}^{-1}\approx 0.37$.
\placefigure{fig5}

The stellar masses of our clean sources range from $1.68\times 10^9$ to $9.77\times 10^{11}\,M_{\odot}$, exhibit a media and $68\%$ dispersion of $\log(M_*/M_{\odot})=10.82_{-0.53}^{+0.41}$ which is the modest range of the main sequence galaxies at similar redshift \citep{Speagle2014}. The age of galaxy plotted in Figure \ref{fig5} is not the time since the onset of star formation which is widely used in simple stellar population (SSP) models but has no real physical meaning when the continuous SFH is applied (\citealt{daCunha2015}). In this work, we use the mass-weighted age, defined as
\begin{equation}
\mathrm{age}_M=\frac{\int_0^t\mathrm{d}t'\,t'\psi(t-t')}{\int_0^t\mathrm{d}t'\psi(t-t')}
\end{equation}
to describe the overall age of the galaxy model, where $\psi(t-t')$ is the star formation history of each model. For our clean sample, the distribution of this age has a median of 2.09\,Gyr, while the central 68th percentile range is 1.12--3.63\,Gyr. As mentioned above, {\sc MAGPHYS} assumes an exponentially declining model with a timescale paramter $\gamma$ to describe the continuous SFH. The resulting distribution of $\gamma$ in Figure \ref{fig5} shows that the median star formation timescale is 3.33\,Gyr ($\gamma=0.30\,\mathrm{Gyr}^{-1}$), larger than the median mass-weighted age. Combining the time when the star formation starts with $\gamma$, we are able to compute the ratio between the present instantaneous SFR $\psi$ and the initial SFR $\psi_0$. This ratio is in the mdeian larger than the e-folding value $\mathrm{e}^{-1}$ by 0.04, suggesting that more than half of our galaxies do not reach their e-folding time of star formation yet. In fact, nearly $90\%$ of our clean sample present a $\psi/\psi_0$ ratio of larger than $10\%$. Therefore, most of these galaxies still maintain a fairly active star formation, which lead them to product enough dust as well as become bright in infrared.

\begin{figure}
\centering
\plotone{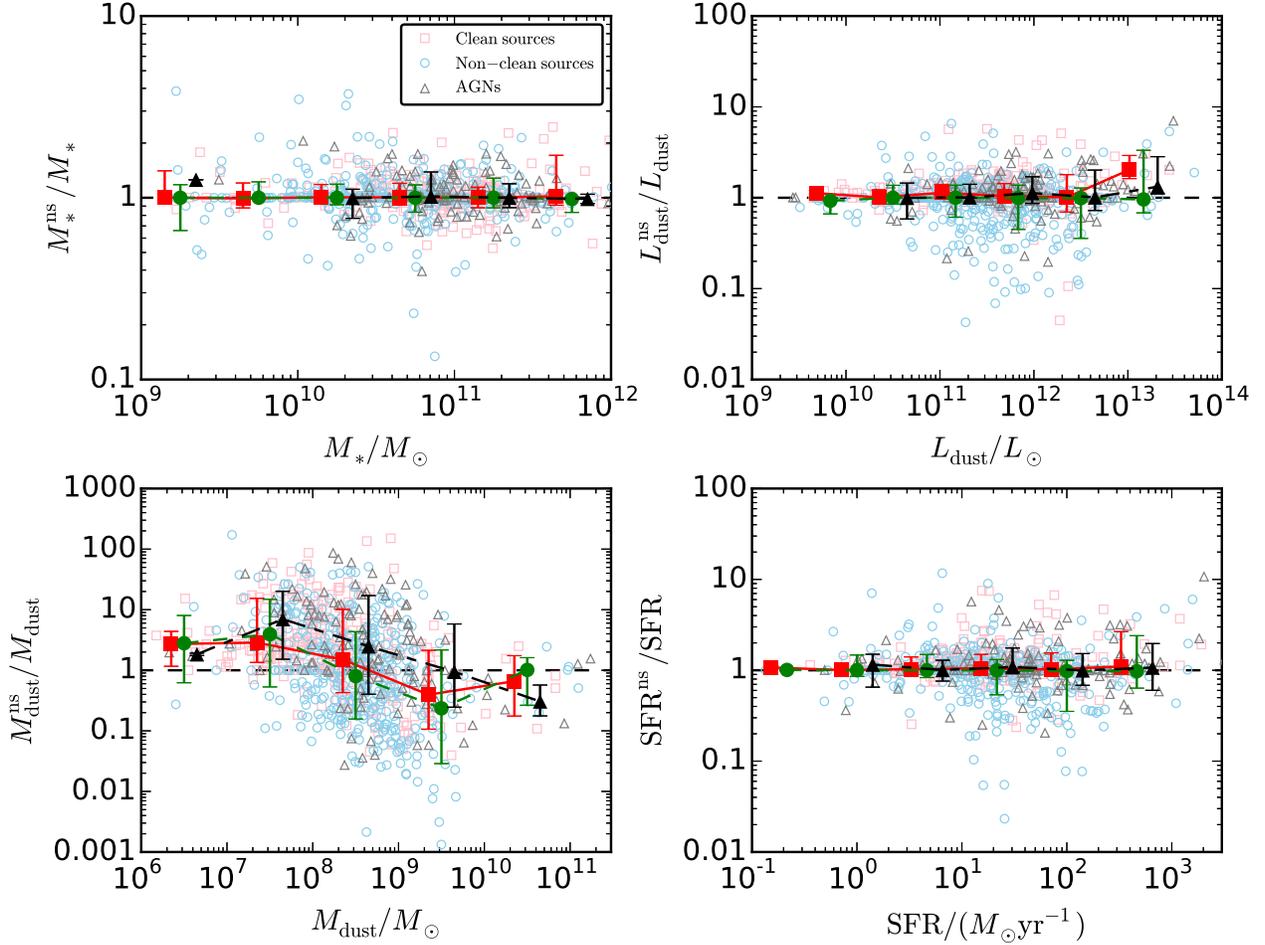}
\caption{Comparison of the stellar mass $M_*$, dust emission $L_{\mathrm{dust}}$, dust mass $M_{\mathrm{dust}}$, star formation rate between SED fitting with and without {\it Herschel}/SPIRE observations. Physical quantities derived from fitting without submillimeter observations are denoted as $Q^{\mathrm{ns}}$ (e.g., $M_*^{\mathrm{ns}}$, $L_{\mathrm{dust}}^{\mathrm{ns}}$). The squares, circles and triangles represent clean source, non-clean sources and AGNs, respectively.}
\label{fig6}
\end{figure}

We include {\it Herschel}/SPIRE observations in our SED fitting, but it is well known that these observations suffered serious confusion noise. To investigate how confusion noise affect the result of SED fitting, we perform a repeated fitting using data without SPIRE measurements (e.g., 250, 350 and 500\,$\mathrm{\mu m}$ bands) and compare some derived quantities, denoted as $Q^{\mathrm{ns}}$, with previous result in Figure \ref{fig6}. Statistically, for stellar mass $M_*$, dust luminosity $L_{\mathrm{dust}}$ and SFR\footnote{SFR from {\sc MAGPHYS} correspond to SFR averaged over the past 100 Myr.}, SED fitting without submillimeter data has nearly no effect on these quantities. That is because $M_*$ is almost determined by optical and NIR observations, and due to the energy balance technique of {\sc MAGPHYS} code, using data only from PACS is enough to constrain $L_{\mathrm{dust}}$ and SFR. However, owing to the fact that $M_{\mathrm{dust}}$ is dominated by large dust grains which is in thermal equilibrium at relatively low temperature, observations from SPIRE are necessary to constrain the cold component of dust emission. As a result, SED fitting without these observations could not give a reliable estimate of $M_{\mathrm{dust}}$. For all clean sources that have SPIRE observations, the $L_{\mathrm{dust}}^{\mathrm{ns}}/L_{\mathrm{dust}}$ ratio has a median value and the 16th-84th percentile range of $1.08_{-0.09}^{+0.33}$. Thus the confusion noise of SPIRE observations has no effect on our results about $L_{\mathrm{dust}}$. An additional SED fitting without using all {\it Herschel} observations is also performed for comparing. We find that only $M_*$ can hold a nearly unchanged result with a little larger dispersion, while $L_{\mathrm{dust}}$, $M_{\mathrm{dust}}$ and SFR can not trace the previous results anymore. The tendencies of the last three quantities are all similar to that of $M_{\mathrm{dust}}$ in Figure \ref{fig6}, present overestimate at low-value end and underestimate at high-value end. Therefore, constraint from FIR observations is necessary to obtain a reliable estimate of $L_{\mathrm{dust}}$ as well as SFR.

In the meantime, we also check the differences of SED fitting results when apply the CB07 version of BC03 models to calculate stellar emission. We find that only stellar mass $M_*$ exhibit a significant systematic underestimate comparing with result of the previous BC03 models, while $L_{\mathrm{dust}}$, $M_{\mathrm{dust}}$ and SFR all present a nearly statistically unchanged result. Consequently, changing the choice of stellar emission models between BC03 and its CB07 version could not change our conclusions on dust luminosity.

Furthermore, the main difference between BC03 and CB07 is the updated treatment of the  thermally pulsing asymptotic giant branch (TP-AGB) phase of stellar evolution, which would lead to a lower stellar mass and a smaller age under the CB07 SSP models \citep{Bruzual2007}. As expected, the mass of stellar populaton $M_*[\mathrm{CB07}]$ is lower than $M_*[\mathrm{BC03}]$ by 25\% in the median which is smaller than the range of 50\%--80\% reported by \citet{Bruzual2007}. On the other hand, the $\mathrm{age}_M[\mathrm{CB07}]/\mathrm{age}_M[\mathrm{BC03}]$ ratio is close to unity with a median value of 0.94 for clean sources. These differences between our result and \citet{Bruzual2007} might be due to two reason: (1) our fitting applied composite stellar population (CSP) models which contain stars with different ages given by SFH introduced above, but not the SSP models used by \citet{Bruzual2007}; (2) the aforementioned mass-weighted age is larger than 2\,Gyr in the median when the differences between BC03 and CB07 models have gone through their maximum and become smaller. In definition, the mass-weighted age reflects the time when most of the stellar mass formed in one CSP model. Thus, this parameter of star-forming galaxies are highly depend on the assumed SFH employed in the SED fitting (\citealt{Conroy2013b}). However, the exponentially declining form of the SFH is unchanged when applied CB07 models and the resulting $\gamma_{\mathrm{CB07}}/\gamma_{\mathrm{BC03}}$ ration has a median of 1.0. Furthermore, for a given SFH form, \citet{Wuyts2011a} found that to strictly constrain $\mathrm{age}_M$ observations at the wavelength of $\lambda_{\mathrm{rest}}<0.2\,\mathrm{\mu m}$, where TP-AGB stars have almost no contribution (\citealt{Bruzual2007}), is required. Therefore, the mass-weighted ages of our clean sample show statistically little change when the CB07 models were used. In the case of stellar mass, the relative old stellar population contribute quite a bit mass of our clean sources, suggesting by their large median $\mathrm{age}_M$, and only a little in the difference between $M_*[\mathrm{CB07}]$ and $M_*[\mathrm{BC03}]$. In combination with younger population, especially those formed in recent 0.1--1\,Gyr \citep{Bruzual2007}, the median $M_*[\mathrm{CB07}]/M_*[\mathrm{BC03}]$ ratio becomes larger than the range given by \citet{Bruzual2007}.

\section{Summary}
\label{sect:summary}
In this work, to investigate the accuracy of monochromatic extrapolated IR luminosity, we utilize multi-wavelength data of GOODS-North and GOODS-South fields to perform a UV-to-submillimeter SED fitting and use the output dust luminosity as a reference value. The main conclusions are as follows:
\begin{enumerate}
\renewcommand{\labelenumi}{(\theenumi)}
\item We compare extrapolated $L_{\mathrm{IR}}^{24}$ and $L_{\mathrm{IR}}^{\mathrm{PACS}}$ based on different templates and conclude that CE01 and DH02 templates present nearly the same estimate in these two bands, while W08 template show large differences and should be used with caution for galaxies at $z\lesssim 0.5$.
\item For CE01 template, extrapolations from PACS bands can estimate the actual IR luminosity out to $z\sim 3.5$ well. A few high-redshift galaxies hint that this consistency may be hold even at $z\sim 5$. However, extrapolations from MIPS 24\,$\mathrm{\mu m}$ for galaxies with $z>1.5$ result in a serious overestimate of the total IR luminosities. $L_{\mathrm{IR}}^{\lambda}$ from SPIRE bands also could not provide reliable estimate of $L_{\mathrm{dust}}$. For DH02 template, the $L_{\mathrm{IR}}^{\lambda}/L_{\mathrm{dust}}$ ratios from 24--160\,$\mathrm{\mu m}$ bands are similar to that of CE01 template, but show a little redshift evolution in 70\,\um\ band. For clean source at $z\lesssim 2$, extrapolations from SPIRE bands present a rough unbiased esitmate of $L_{\mathrm{dust}}$. In the case of W08 template, the absence of a significant ``mid-IR excess'' problem makes it useful to derive $L_{\mathrm{IR}}$ in MIPS 24\,\um\ band, while extrapolations from submillimeter bands also can provide unbiased estimate for galaxies at $z<2$ as DH02.
\item Among the three templates we concerned, the CE01 template provides the best estimate of $L_{\mathrm{dust}}$  in PACS bands, while the DH02 and W08 templates perform better in SPIRE bands although the dispersion is still large. For extrapolations from MIPS 24\,\um\ band, only W08 template does not suffer a significant mid-IR excess problem.
\item To obtain a reliable estimate of the actual IR luminosity using the monochromatic extrapolation method described in this work, our suggestions are as follow. To extrapolate from MIR bands (e.g., MIPS 24\,\um), CE01 template should be used for galaxies at $z<1.5$, and W08 template should be used for galaxies at $z>1.5$ although it will lead to an overestimate of nearly 90\%.  Using FIR observations (e.g., PACS 70--160\,\um) to do this, CE01 template is the best choice out to $z\sim3.5$. Besides, if only submillimeter bands (e.g., SPIRE 250--500\,\um) are available, both DH02 and W08 template can be used for galaxies at $z\lesssim 2$, but the large uncertainty also should be kept in mind. Moreover, $L_{\mathrm{IR}}^{\mathrm{PACS}}$, which is derived from the available longest wavelength PACS band, based on CE01 template can be a good estimator.
\item All the three templates exhibit different degrees of enhanced emission at $\lambda_{\mathrm{rest}}\lesssim 10\,\mathrm{\mu m}$, but well describe the emission of 10--100\,\um\ range of the IR SED. Only DH02 template show a nearly unbiased estimate of the emission of the rest-frame submillimeter part.
\end{enumerate}

\acknowledgments

This work is based on observations taken by the 3D-HST Treasury Program (GO 12177 and 12328) and CANDELS Multi-Cycle Treasury Program with the NASA/ESA {\it HST}, which is operated by the Association of Universities for Research in Astronomy, Inc., under NASA contract NAS5-26555.
This research has made use of data from HerMES project (http://hermes.sussex.ac.uk/). HerMES is a Herschel Key Programme utilising Guaranteed Time from the SPIRE instrument team, ESAC scientists and a mission scientist. The HerMES data was accessed through the Herschel Database in Marseille (HeDaM - http://hedam.lam.fr) operated by CeSAM and hosted by the Laboratoire d'Astrophysique de Marseille.
This work is supported by the National Natural Science Foundation of China (NSFC, Nos. 11673004, 11303002, 11225315, 1320101002, 11433005, and 11421303), the Specialized Research Fund for the Doctoral Program of Higher Education (SRFDP, No. 20123402110037), the Strategic Priority Research Program ``The Emergence of Cosmological Structures'' of the Chinese Academy of Sciences (No. XDB09000000), the Chinese National 973 Fundamental Science Programs (973 program) (2015CB857004), and the Yunnan Applied Basic Research Projects (2014FB155).

\bibliography{draft-aj}
%\bibliography{../IRtemp}

\clearpage

\begin{deluxetable}{cccc}
\tablewidth{0pt}
\tabletypesize{\footnotesize}
%\scriptsize
\tablecaption{Filters adopted in multi-wavelength SED fitting \label{tab1}}
\tablehead{
\multicolumn{2}{c}{GOODS-North} & \multicolumn{2}{c}{GOODS-South} \\
\cmidrule(lr){1-2} \cmidrule(lr){3-4}
\colhead{Filters} & \colhead{Telescope/Instrument} & \colhead{Filters} & \colhead{Telescope/Instrument}
}
\startdata
{\it U} & KPNO 4\,m/Mosaic & {\it U, R} & VLT/VIMOS \\
{\it F{\rm 435}W, F{\rm 606}W, F{\rm 775}W, F{\rm 850}LP} & {\it HST}/ACS & {\it U{\rm 38}, B, V, $\mathit{R_c}$, I} & WFI 2.2\,m \\
{\it B, V, $\mathit{R_c}$, $\mathit{I_c}$, z$^\prime$} & Subaru/Suprime-Cam & IA427, IA505, IA527, IA574,  &  \multirow{2}*{Subaru/Suprime-Cam} \\
{\it F{\rm 125}W, F{\rm 140}W, F{\rm 160}W} & {\it HST}/WFC3 & IA624, IA679, IA738, IA767 & \\
{\it J, H, $\mathit{K_s}$} & Subaru/MOIRCS & {\it F{\rm 435}W, F{\rm 606}W, F{\rm 775}W, F{\rm 850}LP} & {\it HST}/ACS\tablenotemark{a} \\
3.6, 4.5, 5.8, 8\,$\mathrm{\mu m}$ & {\it Spitzer}/IRAC & {\it F{\rm 606}W, F{\rm 814}W, F{\rm 850}LP} & {\it HST}/ACS\tablenotemark{b} \\
24\,$\mathrm{\mu m}$ & {\it Spitzer}/MIPS & {\it F{\rm 125}W, F{\rm 140}W, F{\rm 160}W} & {\it HST}/WFC3 \\
100, 160\,$\mathrm{\mu m}$ & {\it Herschel}/PACS & {\it J, H, $\mathit{K_s}$} & VLT/ISAAC \\
250, 350, 500\,$\mathrm{\mu m}$ & {\it Herschel}/SPIRE & {\it J, $K_s$} & CFHT/WIRcam \\
 & & 3.6, 4.5, 5.8, 8\,$\mathrm{\mu m}$ & {\it Spitzer}/IRAC \\
 & & 24$\,\mathrm{\mu m}$ & {\it Spitzer}/MIPS \\
 & & 70, 100, 160$\,\mathrm{\mu m}$ & {\it Herschel}/PACS \\
 & & 250, 350, 500$\,\mathrm{\mu m}$ & {\it Herschel}/SPIRE \\
\enddata

\tablenotetext{a}{From GOODS (\citealt{Giavalisco2004}).}
\tablenotetext{b}{From CANDELS (\citealt{Grogin2011,Koekemoer2011}).}

\end{deluxetable}

\clearpage

\begin{deluxetable}{ccccccc}
\tablewidth{0pt}
\tablecaption{Statistical description of the $L_{\mathrm{IR}}^{\lambda}/L_{\mathrm{dust}}$ ratio for clean sources in Figure \ref{fig3} \label{tab2}}
\tablehead{
\colhead{\multirow{2}{*}{$\lambda_{\mathrm{obs}}$}} & \multicolumn{3}{c}{$z<1.5$} & \multicolumn{3}{c}{$z\geq 1.5$} \\
\cmidrule(lr){2-4} \cmidrule(lr){5-7}
 & \colhead{CE01} & \colhead{DH02} & \colhead{W08} &  \colhead{CE01} & \colhead{DH02} & \colhead{W08}
}
\startdata
24\,\um & $1.05_{-0.32}^{+0.38}$ & $1.16_{-0.37}^{+0.78}$ & $0.99_{-0.31}^{+0.51}$ & $4.65_{-2.13}^{+3.79}$ & $3.94_{-1.96}^{+3.44}$ & $1.86_{-0.79}^{+1.61}$ \\
70\,\um & $0.92_{-0.24}^{+0.34}$ & $0.81_{-0.18}^{+0.26}$ & $0.76_{-0.18}^{+0.32}$ & $1.31_{-0.74}^{+0.73}$ & $1.48_{-0.85}^{+0.76}$ & $1.87_{-1.08}^{+0.99}$ \\
100\,\um & $0.89_{-0.19}^{+0.17}$ & $0.92_{-0.21}^{+0.17}$ & $0.97_{-0.23}^{+0.20}$ & $1.15_{-0.31}^{+0.44}$ & $1.05_{-0.28}^{+0.36}$ & $1.30_{-0.40}^{+0.59}$ \\
160\,\um & $1.07_{-0.23}^{+0.33}$ & $1.19_{-0.26}^{+0.27}$ & $1.25_{-0.25}^{+0.30}$ & $1.17_{-0.31}^{+0.44}$ & $1.14_{-0.29}^{+0.33}$ & $1.31_{-0.34}^{+0.50}$ \\
PACS & $1.07_{-0.23}^{+0.33}$ & $1.19_{-0.26}^{+0.27}$ & $1.25_{-0.27}^{+0.30}$ & $1.17_{-0.31}^{+0.43}$ & $1.14_{-0.32}^{+0.32}$ & $1.31_{-0.40}^{+0.48}$ \\
250\,\um & $1.19_{-0.69}^{+0.80}$ & $1.02_{-0.57}^{+0.83}$ & $1.04_{-0.55}^{+0.56}$ & $0.65_{-0.20}^{+0.39}$ & $0.72_{-0.21}^{+0.42}$ & $0.75_{-0.21}^{+0.26}$ \\
350\,\um & $1.55_{-0.69}^{+1.22}$ & $1.20_{-0.62}^{+0.92}$ & $1.10_{-0.49}^{+0.65}$ & $0.90_{-0.47}^{+0.79}$ & $0.96_{-0.54}^{+0.86}$ & $0.81_{-0.39}^{+0.36}$ \\
500\,\um & $1.77_{-1.29}^{+2.63}$ & $1.21_{-0.89}^{+1.89}$ & $1.13_{-0.80}^{+1.19}$ & $1.19_{-0.71}^{+0.86}$ & $1.09_{-0.64}^{+0.79}$ & $0.81_{-0.42}^{+0.35}$
\enddata
\tablecomments{$\lambda_{\mathrm{obs}}$ is the observed band that used to derived the template-based monochromatic extrapolated IR luminosity $L_{\mathrm{IR}}^{\lambda}$. The value with error shows the median and the 16th--84th percentile range for each case. }
\end{deluxetable}

\clearpage

\begin{deluxetable}{ccccccccc}
\tablewidth{0pt}
\tablecaption{Statistical description of the $L_{\mathrm{IR}}^{\lambda}/L_{\mathrm{dust}}$ ratio for clean sources in Figure \ref{fig4} \label{tab3}}
\tablehead{
\colhead{$\lambda_{\mathrm{rest}}$(\um)} & \colhead{6.67} & \colhead{11.9} & \colhead{21.1} &  \colhead{37.5} & \colhead{66.7} & \colhead{119} & \colhead{211} & \colhead{375}
}
\startdata
CE01 & $5.20_{-2.47}^{+3.98}$ & $1.14_{-0.35}^{+0.49}$ & $0.94_{-0.26}^{+0.31}$ & $1.01_{-0.33}^{+0.38}$ & $0.92_{-0.22}^{+0.27}$ & $1.08_{-0.39}^{+0.50}$ & $1.56_{-0.95}^{+1.29}$ & $1.65_{-1.15}^{+3.24}$ \\
DH02 & $3.84_{-1.82}^{+3.74}$ & $1.48_{-0.55}^{+0.81}$ & $0.93_{-0.23}^{+0.29}$ & $0.89_{-0.24}^{+0.33}$ & $1.00_{-0.26}^{+0.28}$ & $1.14_{-0.41}^{+0.44}$ & $1.18_{-0.73}^{+1.04}$ & $1.05_{-0.72}^{+2.07}$ \\
W08 & $1.85_{-0.79}^{+1.86}$ & $1.19_{-0.39}^{+0.53}$ & $0.81_{-0.24}^{+0.46}$ & $0.99_{-0.32}^{+0.45}$ & $1.01_{-0.31}^{+0.28}$ & $1.10_{-0.43}^{+0.38}$ & $1.10_{-0.60}^{+0.66}$ & $1.33_{-0.79}^{+2.10}$
\enddata
\tablecomments{The rest-frame wavelength $\lambda_{\mathrm{rest}}$ is the de-redshifted wavelength of $\lambda_{\mathrm{obs}}$ that used to derived $L_{\mathrm{IR}}^{\lambda}$.  The whole 5--500\,$\mathrm{\mu m}$ wavelength range was separated into 8 bins in logarithmic space. The value with error shows the median and the 16th--84th percentile range for each case.}
\end{deluxetable}

\end{document}